\documentclass[aps,pre,superscriptaddress,twocolumn,showpacs]{revtex4}
\usepackage{epsfig}
\usepackage{graphicx}
\usepackage{dcolumn}
\usepackage{bm}
\usepackage{amsmath,amssymb,times}

\begin{document}

\title{Suppressing traffic-driven epidemic spreading by edge-removal strategies}

\author{Han-Xin Yang}\email{hxyang01@gmail.com}
\affiliation{Department of Physics, Fuzhou University, Fuzhou
350108, China}

\author{Zhi-Xi Wu}
\affiliation{Institute of Computational Physics and Complex Systems,
Lanzhou University, Lanzhou, Gansu 730000, China}

\author{Bing-Hong Wang}
\affiliation{Department of Modern Physics, University of Science and
Technology of China, Hefei 230026, China}

\begin{abstract}

The interplay between traffic dynamics and epidemic spreading on
complex networks has received increasing attention in recent years.
However, the control of traffic-driven epidemic spreading remains to
be a challenging problem. In this Brief Report, we propose a method
to suppress traffic-driven epidemic outbreak by properly removing
some edges in a network. We find that the epidemic threshold can be
enhanced by the targeted cutting of links among large-degree nodes
or edges with the largest algorithmic betweeness. In contrast, the
epidemic threshold will be reduced by the random edge removal. These
findings are robust with respect to traffic-flow conditions, network
structures and routing strategies. Moreover, we find that the
shutdown of targeted edges can effectively release traffic load
passing through large-degree nodes, rendering a relatively low
probability of infection to these nodes.

\end{abstract}

\date{\today}


\maketitle

Human society has always suffered from various viruses, such as
AIDS, H1N1 influenza and computer virus. As the rapid development of
complex network theory~\cite{network1,network2}, much effort has
been dedicated to understand dynamical processes of epidemic
spreading on complex networks in the past
decade~\cite{epidemic1,epidemic2,
epidemic3,epidemic4,epidemic5,epidemic6,epidemic7,epidemic8,epidemic9}.
Propagation is usually assumed to be driven by reaction processes,
in the sense that every infected node transmits diseases to all its
neighbors at each time step, producing a diffusion of the epidemics
in the population. However, in many realistic situations, even when
there is a link connecting two nodes, infection will not propagate
unless some kind of traffic happens between the nodes. For example,
a computer virus can spread over Internet via email-exchanges. In
the absence of such data packet transmission, even if there is a
path linking two computers, an infected computer will not be able to
infect the other one. Another example is that air transport
tremendously accelerates the propagation of infectious diseases
among different countries.

The first attempt to incorporate traffic into epidemic spreading is
based on metapopulation model~\cite{1,2,3,4,5,6,7,8}. This framework
describes a set of spatially structured interacting subpopulations
as a network, whose links denote the traveling path of individuals
across subpopulations. Each subpopulation consists of a large number of individuals. Recently, Meloni $et$ $al$. proposed another
traffic-driven epidemic spreading model~\cite{Meloni}, in which each
node of a network represents a router and the epidemic can spread
between nodes by the transport of information packets.

One of the most important issues in the study of epidemic spreading
is how to control the prevalence of infection. To suppress the
traffic-driven epidemic spreading, a variety of strategies have been
considered, such as the restriction of traffic flow~\cite{flow}, the
selection of routings~\cite{yang} and heterogeneous curing
rate~\cite{hou}, etc. In this Brief Report, we propose a method to
control traffic-driven epidemic spreading based on edge-removal
strategies. The principle of edge-removal strategies is to affect
the spreading dynamics of epidemics (or virus) by deleting some
edges in the underlying network. It has been recognized that
edge-removal strategies can greatly influence the dynamics of
synchronization~\cite{yin}, evolutionary games~\cite{jiang} and
traffic~\cite{hu}. In Ref.~\cite{zhang}, Zhang $et$ $al$. found
that, both random and targeted deletion of edges can suppress the
outbreak of reaction-based epidemic. In contrast, we will show that
random and targeted edge-removal strategies play different roles in
the traffic-driven epidemic spreading. Specifically, we have found
that the random shutdown of edges decreases the epidemic threshold,
while the targeted shutdown of edges increases the epidemic
threshold.

Following the work of Meloni $et$ $al$.~\cite{Meloni}, we
incorporate the traffic dynamics into the
susceptible-infected-susceptible model~\cite{SIS} of epidemic
spreading as follows. In a network of size $N$, at each time step,
$\lambda N$ new packets are generated with randomly chosen sources
and destinations, and each node $i$ can deliver at most $C_{i}$
packets toward their destinations. Packets are forwarded according
to a given routing algorithm. The queue length of each agent is
assumed to be unlimited. The first-in-first-out principle applies to
the queue. Each newly generated packet is placed at the end of the
queue of its source node. Once a packet reaches its destination, it
is removed from the system. Nodes can be in two discrete states,
either susceptible or infected. After a transient time, the total
number of delivered packets at each time will reach a steady value.
Subsequently, an initial fraction of nodes $\rho_{0}$ is set to be
infected (we choose $\rho_{0}=0.1$ in our numerical experiments).
The infection spreads in the network through packet exchanges. All
packets queuing in an infected node are infected, while all packets
in a susceptible node are uninfected. A susceptible node has the
probability $\beta$ of being infected every time it receives an
infected packet from any infected neighboring nodes. With
probability $1-\beta$, the virus in an infected packet will be
cleaned by antivirus software in the susceptible node. The infected
nodes are recovered at rate $\mu$ (here, we set $\mu=1$).

In the following, we carry out simulations systematically by
employing traffic-driven epidemic spreading on the Barab\'{a}si-Albert (BA) scale-free networks~\cite{BA} with the
shortest-path routing algorithm~\cite{shortest1,shortest2}.
Initially, the size of BA network is set to be $N=5000$ and the
average degree of the network $\langle k \rangle=10$. Moreover, we
assume that the node delivering capacity is infinite, so that
traffic congestion will not occur in the network.

Three edge-removal strategies are considered respectively. (I) The
random strategy (RS): We randomly remove a fraction $f$ of edges
from the network. (II) The betweenness-based strategy (BS): We rank
the edges in descending order according to their algorithmic
betweenness. The algorithmic betweenness of an edge is the average
number of packets passing through that edge at each time step in the
steady state. We close a proportion of edges at the top of the
ranking list. (III) The degree-based strategy (DS): We define the
significance $G_{ij}$ of an edge by the product of the degrees of
two nodes $i$ and $j$ at both sides of the edge, i.e.,
$G_{ij}=k_{i}\times k_{j}$. After computing the significance of all
edges, we rank the edges in descending order according to their
significance. A proportion of edges at the top of the ranking list
are removed from the network. For all three strategies, disconnected
networks are avoided.

\begin{figure}
\begin{center}
 \scalebox{0.4}[0.4]{\includegraphics{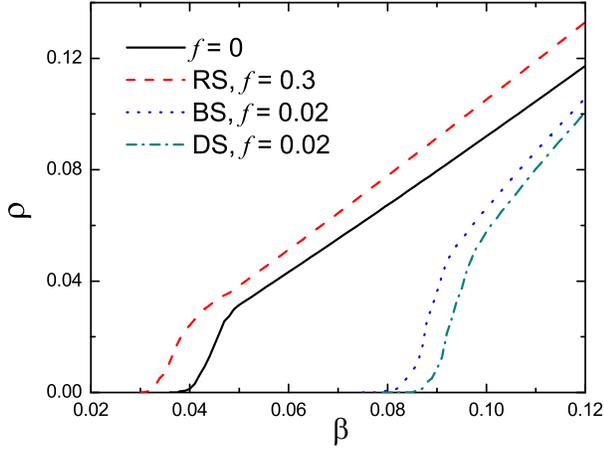}} \caption{(Color online) Density of infected nodes $\rho$ as a function of the spreading rate
$\beta$ for the null ($f=0$), RS ($f=0.3$), BS ($f=0.02$) and DS
($f=0.02$) cases. The packet-generation rate $\lambda=1$. Each curve
is an average of $10^{2}$ different realizations.}
 \label{1}
\end{center}
\end{figure}

Figure~\ref{1} shows the density of infected nodes $\rho$ as a
function of the spreading rate $\beta$ for the null ($f=0$, i.e., no
edges are shutdown during the epidemic spreading process), RS, BS
and DS cases. We observe that for each case, there exists an
epidemic threshold $\beta_{c}$, beyond which the density of infected
nodes is nonzero and increases as $\beta$ is increased. For
$\beta<\beta_{c}$, the epidemic goes extinct and $\rho=0$.

\begin{figure}
\begin{center}
\scalebox{0.4}[0.4]{\includegraphics{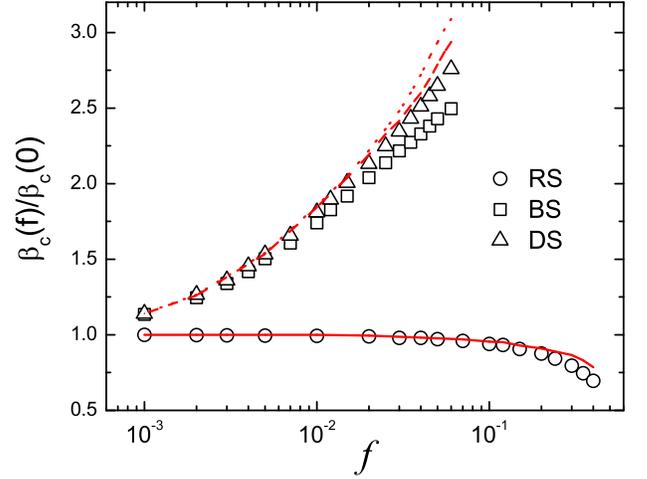}} \caption{(Color
online) The relative ratio $\beta_{c}(f)/\beta_{c}(0)$ as a function
of the fraction of deleted edges $f$ for RS, BS and DS. The
packet-generation rate $\lambda=1$ and $\beta_{c}(0)\simeq 0.043$.
Each data point results from an average over $10^{2}$ different
realizations. The curves are theoretical predictions according to
Eq. (1) and Eq. (2). The solid, dashed, and dotted curves correspond
to the theoretical predictions for RS, BS and DS, respectively.}
 \label{2}
\end{center}
\end{figure}

Figure~\ref{2} shows the ratio of $\beta_{c}(f)$ to $\beta_{c}(0)$
as a function of the fraction of deleted edges $f$ for the cases of RS, BS and DS. Here $\beta_{c}(0)$ is the epidemic threshold for the null case and $\beta_{c}(f)$ represents the epidemic threshold under
the condition that a fraction $f$ of edges in the network are
deleted. From Fig.~\ref{2}, we can see that
$\beta_{c}(f)/\beta_{c}(0)>1$ and $\beta_{c}(f)/\beta_{c}(0)$
increases with the increment of $f$ for the cases of BS and DS, indicating that targeted edge-removal strategies can effectively suppress the outbreak of epidemic. As shown in Fig.~\ref{2}, compared with that of the null case, the epidemic threshold can be enhanced more than 50\% when only one percent of targeted edges are cutting-down. It is also noted that the epidemic threshold in the case of DS is a little larger than that in the case of BS, given that the same fraction of edges are deleted. For RS, however, $\beta_{c}(f)/\beta_{c}(0)$ is found to decrease as $f$ increases, demonstrating that the random edge-removal strategy is failed to inhibit the spreading of epidemic, but rather enhances its propagation.

According to the analysis of Ref.~\cite{Meloni}, the epidemic
threshold for uncorrelated networks is
\begin{equation}
\beta_{c}=\frac{\langle b_{\mathrm{alg}} \rangle}{\langle b_{\mathrm{alg}}^{2}
\rangle}\frac{1}{\lambda N},
\end{equation}
where $b_{\mathrm{alg}}$ is the algorithmic betweenness of a
node~\cite{alg1,alg2} and $\langle \cdot \rangle$ denotes the
average of all nodes. The algorithmic betweenness of a node is the
number of packets passing through that node when the
packet-generation rate $\lambda=1/N$~\cite{alg1,alg2}. For the shortest-path routing protocol, the algorithmic betweenness is equal to the topological betweenness ($b_{\mathrm{alg}}=b_{\mathrm{top}}$) and $\langle b_{\mathrm{alg}} \rangle=\langle
D \rangle/(N-1)$, where $\langle D \rangle$ is the average
topological distance of a network. Here, the topological betweenness of a node $k$ is defined as
\begin{equation}
b_{\mathrm{top}}^{k}=\frac{1}{N(N-1)}\sum_{i\neq
j}\frac{\sigma_{ij}(k)}{\sigma_{ij}},
\end{equation}
where $\sigma_{ij}$ is the total number of shortest paths going from $i$ to $j$, and $\sigma_{ij}(k)$ is the number of shortest paths going from $i$ to $j$ and passing through $k$. The average topological distance of a network is given by $\langle D \rangle=\sum_{i\neq j}d_{ij}/[N(N-1)]$, where $d_{ij}$ is the shortest distance between $i$ and $j$. Combining Eq.~(1) and Eq.~(2), we are able to calculate the theoretical values of $\beta_{c}(f)/\beta_{c}(0)$. In Fig.~\ref{2}, we notice that the theoretical predictions agree well with the numerical results.

\begin{figure}
\begin{center}
 \scalebox{0.4}[0.4]{\includegraphics{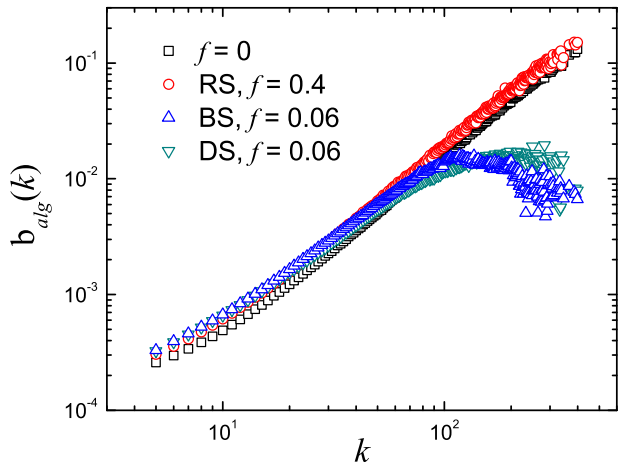}} \caption{(Color online) The algorithmic betweenness $b_{\mathrm{alg}}(k)$ as a function of the original degree $k$ for the null ($f=0$), RS ($f=0.4$), BS ($f=0.06$) and DS ($f=0.06$) cases. Each data point results from an average over $10^{2}$ different realizations.}
 \label{3}
\end{center}
\end{figure}

\begin{figure}
\begin{center}
 \scalebox{0.4}[0.4]{\includegraphics{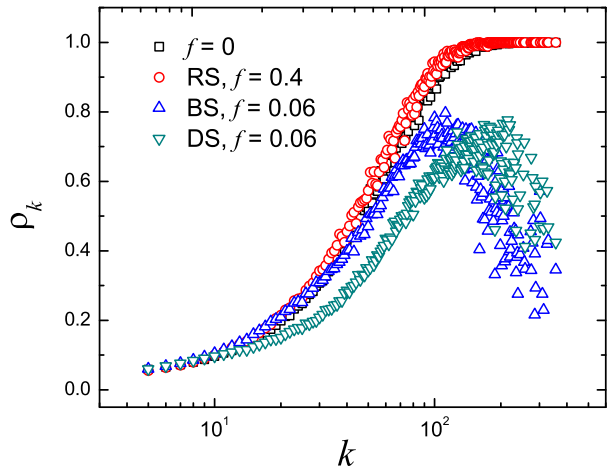}} \caption{(Color online) The dependence of the density of infected nodes $\rho_{k}$ on
 the original degree $k$ for the null ($f=0$), RS ($f=0.4$),
BS ($f=0.06$) and DS ($f=0.06$) cases. In all cases, the
packet-generation rate $\lambda=1$ and the density of infected nodes
$\rho\simeq0.1$. Each data point results from an average over
$10^{2}$ different realizations.}
 \label{4}
\end{center}
\end{figure}

To show how different edge-removal strategies affect traffic flow
on the nodes with different degrees, we display in Fig.~\ref{3} the dependence of algorithmic betweenness $b_{\mathrm{alg}}(k)$ on degree $k$. Here the degree of a node is calculated before the implementation of deleting edges. From Fig.~\ref{3}, one can see that for both the null and RS cases, $b_{\mathrm{alg}}(k)$ increases as the increasing of $k$, and the relationship between $b_{\mathrm{alg}}(k)$ and $k$ follows a power-law form as $b_{\mathrm{alg}}(k) \sim k^{\nu}$. The exponent $\nu$ is almost the same for the null and RS cases. In addition, we can also
observe that, for large values of $k$, $b_{\mathrm{alg}}(k)$ is much smaller in the cases of BS and DS as compared to that in the cases of null and RS. This point is understandable, since the targeted deletion of edges makes many transport paths bypass large-degree nodes and reroute via moderate-degree nodes, hence decreasing the algorithmic betweenness of those hub nodes. Consequently, as shown in Fig.~\ref{3}, the highest values of $b_{\mathrm{alg}}(k)$ are refered to those medium-degree nodes in the cases of BS and DS.

We define $\rho_{k}$ as the density of infected nodes of degree $k$.
Figure~\ref{4} features the dependence of $\rho_{k}$ on $k$ for the
null, RS, BS and DS cases. Combining Figs.~\ref{3} and
~\ref{4}, we can observe that the algorithmic betweenness is
positively correlated with the risk of being infected. As shown in
Fig.~\ref{4}, $\rho_{k}$ increases as $k$ increases for the null and RS cases. Compared with these two cases, the probability of being infected for large-degree nodes is greatly reduced in the cases of BS and DS.

\begin{figure}
\begin{center}
 \scalebox{0.4}[0.4]{\includegraphics{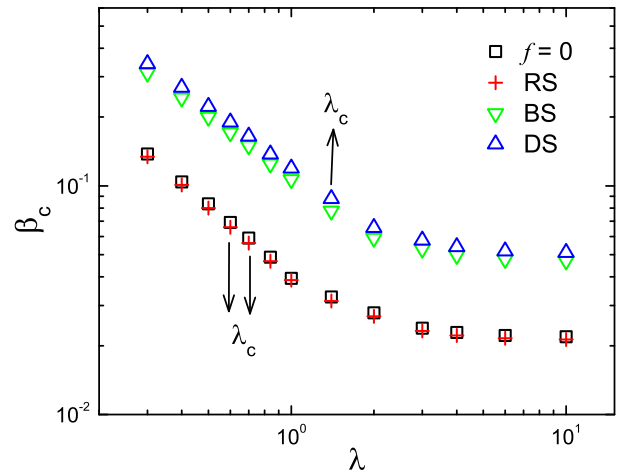}} \caption{(Color online) The epidemic threshold $\beta_{c}$ as a
function of the packet-generation rate $\lambda$ for the null case
($f=0$), the RS case ($f=0.06$), the BS case ($f=0.06$) and the DS case ($f=0.06$). For all cases, the delivery capacity of a node $i$ is equal to its original degree, that is $C_{i}=k_{i}$. The arrows mark the critical packet-generating rates $\lambda_{c}$. For the null case ($f=0$), $\lambda_{c}\approx 0.7$; For the RS case ($f=0.06$),
$\lambda_{c}\approx 0.6$; For the BS and DS cases ($f=0.06$),
$\lambda_{c}\approx 1.4$. Each data point results from an average
over $10^{2}$ different realizations.}
 \label{5}
\end{center}
\end{figure}

We now turn our attention to a more realistic situation where the
node delivering capacity is finite. The main difference with the
infinite-capacity case is the possibility of the emergence of
traffic congestion in the network, which occurs when the
packet-generating rate exceeds a critical value
$\lambda_{c}$~\cite{alg2}. Specially, we set the delivery capacity
of a node $i$ to be equal to its original degree, that is
$C_{i}=k_{i}$. The epidemic threshold $\beta_{c}$ as a function of
the packet-generation rate $\lambda$ for the null, RS, BS and DS
cases are depicted in Fig.~\ref{5}. We see that $\beta_{c}$
decreases and stabilizes at a constant value as $\lambda$ increases.
We also observe that, the random-deletion strategy cannot increases
the epidemic threshold while the targeted-deletion strategies (BS
and DS) can effectively enhance the epidemic threshold, regardless
of that the traffic is in the free-flow state ($\lambda <
\lambda_{c}$) or in the congested state ($\lambda > \lambda_{c}$).

\begin{figure}
\begin{center}
 \scalebox{0.4}[0.4]{\includegraphics{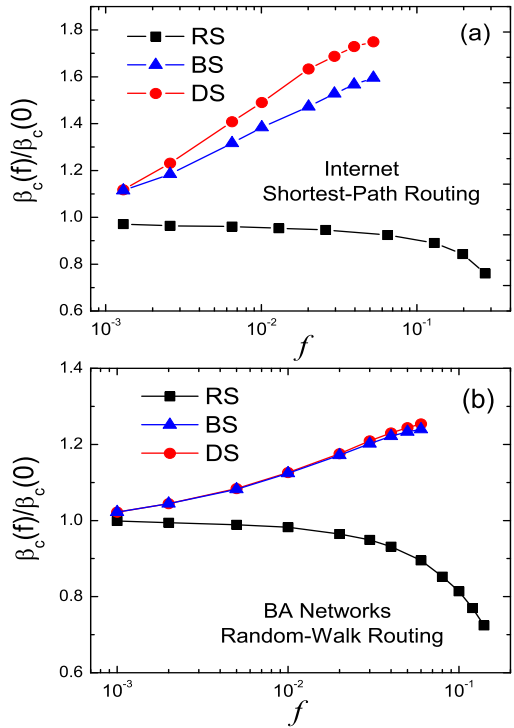}} \caption{(Color online) (a) The relative ratio $\beta_{c}(f)/\beta_{c}(0)$ as a function of the
fraction of deleted edges $f$ for the RS, BS and DS cases. The
Internet at the autonomous system level and the shortest-path
routing are applied. The packet-generation rate $\lambda=0.25$ and
$\beta_{c}(0)\simeq 0.045$. (b) The relative ratio
$\beta_{c}(f)/\beta_{c}(0)$ as a function of the fraction of deleted
edges $f$ for the RS, BS and DS cases. The BA networks and the
random-walk routing are used. The packet-generation rate
$\lambda=0.02$ and $\beta_{c}(0)\simeq 0.058$. For both (a) and (b),
the node delivering capacity is infinite. Each data point results
from an average over $10^{2}$ different realizations.}
 \label{6}
\end{center}
\end{figure}

Finally, we examine the performance of our proposed strategies in controlling epidemic spreading by considering our model on different network structures and with alternative routing protocols. In Fig.~\ref{6}(a), we present the simulation results for different edge-removal strategies on the Internet at the autonomous system level~\cite{data}, where the network size $N = 6474$ and the average degree $\langle k\rangle=3.88$ before cutting edges. The packets are delivered following the shortest-path routing. In Fig.~\ref{6}(b), we carry out our studies on the BA networks, where packets are forwarded according to a random-walk routing algorithm, i.e., a packet is
delivered to a randomly selected neighbor. For the random-walk routing, the algorithmic betweenness $b_{\mathrm{alg}}$ of a node is proportional to its degree~\cite{random1,random2}. As shown in Fig.~\ref{6}, the conclusion that the suppression of epidemic outbreak by the targeted edge-removal strategies and the promotion of epidemic outbreak by the random edge-removal strategy, is still valid.

In conclusion, we have studied the impact of edge-removal strategies
on traffic-driven epidemic spreading. The shutdown of links in terms
of their algorithmic betweeness or of links connecting large degree
nodes, are found to be quite efficient in suppressing epidemic
spreading. Contrary to previous studies on reaction-based
epidemic~\cite{zhang}, we find that the random shutdown of edges
accelerates the outbreak of traffic-driven epidemic. Furthermore,
compared to the deletion of edges with the largest algorithmic
betweeness, the shutdown of links connecting large-degree nodes is
proved to be more effective in enhancing the epidemic threshold.
Thus, according to our present studies, the targeted link-closing
method can be used to control the spreading of computer virus in the
Internet. For example, we can temporarily close links between
large-degree nodes at the time of virus outbreak, and recover these
links after virus is eliminated from the system, which could be
realized readily by special softwares.

\begin{acknowledgments}

H.X.Y. is supported by the Foundation of Fuzhou University under
Grant No. 0110-600607. Z.X.W. acknowledges support from the National
Natural Science Foundation of China (Grant Nos. 11005051, 11135001
and 11147605). B.H.W. is supported by the National Natural Science
Foundation of China (Grant Nos. 11275186 and 91024026).

\end{acknowledgments}

\end{document}